\input harvmac
\sequentialequations
\lref\harvey{J. Harvey, ``Magnetic Monopoles, Duality and Supersymmetry'', in 
Trieste HEP Cosmology 1995, hep-th/9603086. }

\lref\ae{P.~C.~Aichelburg, and F.~Embacher,
``Exact superpartners of N=2 supergravity solitons,'' Phys. Rev. D34 (1986) 3006.}





\lref\aeseries{P.~C.~Aichelburg, and F.~Embacher, ``Supergravity Solitons I; General Framework,''
Phys. Rev. D37 (1988) 338; ``Supergravity Solitons II; the Free Case,'' Phys. Rev. D37 (1988) 911; 
``Supergravity Solitons III; the Background Problem,'' Phys.Rev. D37 (1988) 1436; ``Supergravity 
Solitons IV; Effective Soliton Interaction,'' Phys. Rev. D37 (1988) 2132.}

\lref\witten{E.~Witten and D.~Olive, ``Supersymmetry Algebras That Include Topological Charges,''
Phys. Lett. 78B (1978) 97.}

\lref\bogo{E.~B.~Bogomol'nyi, ``The Stability of Classical Solutions,'' Sov. J. Nucl. Phys. 24 (1976) 
389.}

\lref\jerome{J.~P.~Gauntlett, ``Low-Energy Dynamics of N=2 Supersymmetric Monopoles,''
Nucl. Phys. B411 (1994) 443, hep-th/9305068.}

\lref\jeromedual{J.~P.~Gauntlett, ``Duality and Supersymmetric Monopoles,'' 
Lectures given at 33rd Karpacz Winter School of Theoretical Physics: Duality - Strings and
Fields, hep-th/9705025.}

\lref\manton{N.~S.~Manton, ``A Remark on the Scattering of Monopoles,'' Phys. Lett. 110B (1982) 54.}

\lref\mantonlong{N.~S.~Manton, ``Monopole Interactions at Long Range,'' Phys. Lett. 154B (1985) 397.}

\lref\sen{A.~Sen, ``Dyon-Monopole Bound States, Self-Dual Harmonic Forms on the 
Multi-Monopole Moduli Space, and SL(2,Z) Invariance in String Theory,'' Phys. Lett. B329 (1994) 217.}

\lref\bktw{V.~Balasubramanian, D.~Kastor, J.~Traschen and K.~Z.~ Win, ``The Spin of the M2-Brane and Spin-Spin
Interactions via Probe Techniques,'' hep-th/9811037.}

\lref\jackiw{R.~Jackiw and C.~Rebbi, ``Solitons With Fermion Number 1/2,'' Phys. Rev. D13 (1976) 3398.}

\lref\atiyah{M.~F.~Atiyah and N.~J.~ Hitchin, ``Low-Energy Scattering of Nonabelian Monopoles,''
Phys. Lett. 107A (1985) 21.}

\lref\liu{I.~Giannakis and J.~T.~Liu, ``$N=2$ Supersymmetry and Dipole Moments,'' Phys. Rev. D58 (1998) 2509,
hep-th/9711173.}

\lref\osborn{H.~Osborn, ``Electric Dipole Moment for Supersymmetric Monopoles,'' Phys. Lett. 115B (1982) 226.}

\lref\porrati{S.~Ferrara and M.~Porrati, ``Supersymetric Sum Rules on Magnetic Dipole Moments of Arbitrary Spin
Particles,'' Phys. Lett. 288B (1992) 85.}


\Title{\vbox{\baselineskip12pt
\hbox{UMHEP-457}
\hbox{hep-th/9812191}}}
{\vbox{\centerline{\titlerm Electric Dipole Moment of a BPS Monopole}
 }}
\bigskip
\centerline{
David~Kastor\foot{kastor@phast.umass.edu}
and Euy~Soo~Na\foot{esna@phast.umass.edu}  }
\bigskip
\centerline{\it Department of Physics and Astronomy}
\centerline{\it University of Massachusetts}
\centerline{\it Amherst, MA 01003-4525 USA}
\bigskip\bigskip
\centerline{\bf Abstract}
\medskip
Monopole ``superpartner'' solutions are constructed by acting with finite, broken supersymmetry transformations
on a bosonic $N=2$ BPS monopole.  The terms beyond first order in this construction 
represent the backreaction of the the fermionic zero-mode state on the other fields. Because of the quantum 
nature of the fermionic zero-modes, the superpartner solution is necessarily operator valued.
We extract the electric dipole moment
operator and show that it is proportional to the fermion zero-mode angular momentum operator with a
gyroelectric ratio $g=2$. The magnetic quadrupole operator is shown to vanish identically on all states.
We comment on the usefulness of the monopole superpartner solution for a study of the long-range spin dependent
dynamics of BPS monopoles.
\medskip
\Date{December, 1998}
\vfill\eject
\newsec{Introduction}
It is well known that BPS monopoles of $N=2$ Yang-Mills theory are invariant under half the
supersymmetry generators and hence form a 4-dimensional, short representation of the supersymmetry
algebra \witten\foot{See {\it e.g.} \harvey\ for a good review of this subject.}. 
The fact that monopoles in supersymmetric theories can carry non-zero
spin allows for the strong possibility that Montonen-Olive type dualities 
may actually hold in theories with sufficient amounts of supersymmetry (see \jeromedual\ for a review).
Moreover, the spin-dependent interactions of monopoles in $N=4$ Yang-Mills theory are crucial for
the existence of bound states required by duality \sen.

These results motivate gaining as clear as possible an understanding of the
spin-dependent physics of monopoles. In this paper we contribute to this understanding by studying the
long-range fields of the different states in the $N=2$ BPS monopole supermultiplet.
Following  work of Aichelburg and Embacher on $N=2$ BPS black holes \ae, we generate the fields of a monopole
``superpartner'' solution by acting on the bosonic monopole with an arbitrary, finite, broken supersymmetry
transformation. 

From the work of Jackiw and Rebbi \jackiw, we know that the angular momentum of spinning
monopoles is carried by the quantized states of fermionic zero-modes. For a single BPS monopole, the
fermionic zero-modes are generated by {\it infinitesimal} broken supersymmetry transformations. What we get by
acting with a {\it finite} transformation is then the backreaction of the fermionic
zero-modes on the other fields. For example, since the fermionic fields carry electric charge, the fermionic
zero-mode state acts as a source at quadratic order for the electric field. Because of the
quantized nature of the fermionic zero-mode states \jackiw, the fields of the monopole superpartner solution
are necessarily operator valued.

The results we find are interesting in themselves. For example, the operator valued electric dipole
moment is proportional to the angular momentum operator with a gyroelectric ratio $g=2$ and the
magnetic quadrupole moment tensor is found to vanish identically for all spin states. 
We also point
out that our results would have a more substantial use in a study of spin-dependent monopole dynamics, which 
unfortunately must await further technical developments. 

There are two different ways to study the low energy
dynamics of bosonic monopoles. The first, due to Manton \manton, postulates that
at low energies monopoles follow geodesics on the moduli space of static multi-monopole configurations.
For the case of two monopoles, Atiyah and Hitchen \atiyah\ were then able to construct the exact moduli space
metric. The second approach, also due to Manton \mantonlong, employs
a monopole test-particle or probe propagating in the long-range background $U(1)$ fields of another monopole.
This second approach, while less powerful than the first because of its restriction to large separations,
gives a physically intuitive derivation of the Taub-NUT limit of the moduli space metric.

The low energy spin-dependent dynamics of $N=2$ monopoles have also been studied in a moduli space 
approximation \jerome. Monopole bound states in this treatment are related to normalizable
harmonic forms on the multi-monopole 
moduli space. In the $N=4$ case, the explicit construction of such a form on the two
monopole moduli space \sen\ established the existence of a bound state required by
$S$-duality. It seems likely that one could also study the spin-dependent interactions of a pair of monopoles via
probe techniques\foot{The spin-dependent interactions of $N=2$ BPS black holes \aeseries\ and more recently
M2-branes \bktw\ have been studied using probe techniques.}.  A necessary ingredient would be a
$\kappa$-symmetric superparticle Lagrangian for a BPS monopole propagating in the background fields of $N=2$ 
$U(1)$ Yang-Mills theory.
Such a Lagrangian does not seem to exist in the literature at this point.  
The appropriate background fields for studying spin-dependent interactions in the $N=2$ case,
in analogy with the gravitating cases studied in \aeseries,\bktw,  would
be the long-range $U(1)$ fields of the monopole superpartner configurations presented
here.  In the $N=4$ case, for example, 
one could then hope to identify in a more physically intuitive way the
attractive channel leading to the bound state found in \sen. 

\newsec{Monopole Superpartners}
We now turn to the construction
of the BPS monopole superpartner solutions.
We work in $N=2$ Yang-Mills theory with gauge group $SU(2)$. The lagrangian is given by
\eqn\super{\eqalign{
{\cal L}_{N=2}&={\rm Tr}(-{1\over4}F_{\mu\nu}F^{\mu\nu}
-{1\over4}(D_{\mu}P)^2-{1\over2}(D_{\mu}S)^2-{e^2\over2}[S,P]^2 \cr &
+i\bar\psi\gamma^{\mu}D_{\mu}\psi-e\bar\psi[S,\psi] -e\bar\psi\gamma_5[P,\psi]), \cr} }
where all fields are $SU(2)$ Lie algebra valued, {\it e.g.} $S=S^a T^a$, 
$S$ and $P$ are two scalar Higgs fields and $\psi$ is a Dirac fermion.
The nonabelian electric and magnetic field strengths are defined by
$E^{ai}=-F^{a0i}$ and $B^{ai}=-{1\over2}\epsilon^{ijk}F_{jk}^a$.
Gauge symmetry breaking is imposed through the
boundary condition at infinity $\sum_aS^aS^a = v^2$,
which breaks the $SU(2)$ gauge symmetry down to a $U(1)$ subgroup. The space of possible vacuum values for the
Higgs field $S$ is a two-sphere, leading to the existence of magnetic monopole configurations.

The Lagrangian \super\ is invariant, up to a total derivative term, under the global supersymmetry
transformations  
\eqn\susyvariations{\eqalign{
\delta A_{\mu}&=i\bar\alpha\gamma_{\mu}\psi -i\bar\psi\gamma_{\mu}\alpha,\qquad
\delta P=\bar\alpha\gamma_5\psi
-\bar\psi\gamma_5\alpha, \qquad
\delta S=i\bar\alpha\psi
-i\bar\psi\alpha, \cr
\delta\psi&=(\half\gamma^{\mu\nu}F_{\mu\nu}-i\gamma^\mu D_\mu S +\gamma^\mu D_\mu P\gamma_5-i[P,S]\gamma_5)\alpha, \cr }}
where the parameter $\alpha$ is a Grassmann valued Dirac spinor\foot{Our conventions for the 
Minkowski metric are ``mostly minus''
$\eta_{\mu\nu}={\rm diag}(+1,-1,-1,-1)$ and $\gamma_5=+i\gamma_0\gamma_1\gamma_2\gamma_3$.}. 
For a static, BPS monopole field configuration with $P=A_0=\psi=0$ and 
\eqn\bps{D_iS^a =\half\epsilon_{ijk}F_{jk}^a,}
only the fermion $\psi$ has a nontrivial supersymmetry variation given by
\eqn\fermionvar{\delta\psi=-2(\gamma^kD_kS)P_-\alpha,}
where $P_\pm=\half(1\pm\Gamma_5)$ are projection operators with $\Gamma_5=-i\gamma_0\gamma_5$.
If we define projected spinors $\alpha_\pm$ satisfying $P_\pm\alpha_\pm=\alpha_\pm$, then $\alpha_+$
generates unbroken supersymmetry transformations, while $\alpha_-$ generates broken supersymmetry transformations.
The variation $\delta\psi$ under a broken supersymmetry transformation gives a zero-mode of the fermion field
equation in the monopole background.

Following work of Aichelburg and Embacher \ae\ on $N=2$ BPS black holes, we now look at the
field configuration generated by acting with an arbitrary finite broken symmetry transformation
$\alpha=P_-\alpha$ on a purely bosonic BPS monopole. The finite transformation is obtained by simply iterating
the infinitesimal transformations. Schematically representing all the fields by $\Phi$ and the  original
bosonic field configuration by $\bar\Phi$, we have the expansion
\eqn\taylor{\Phi= e^\delta\bar\Phi
=\bar\Phi +\delta\bar\Phi+{1\over 2}\delta^2\bar\Phi+{1\over 3!}\delta^3\bar\Phi
+{1\over 4!}\delta^4\bar\Phi , }
where, as in \ae, the expansion truncates at fourth order because of the Grassmann nature of $\alpha$. 
The expansion \taylor\ generates an exact solution to the field equations which is non-linear in the broken
supersymmetry parameter $\alpha$. Since the linear term in $\alpha$ \fermionvar\ simply gives the fermion
zero-modes, the full expansion represents the backreaction of these modes on the other fields. 
Following the terminology of \ae, we call this the monopole ``superpartner'' solution. 

As discussed in \bktw\ an interpretational issue arises because the spacetime fields $S$, $P$, $\psi$ and
$A_\mu$  of the superpartner solution appear to be Grassmann valued. The resolution is to recall the work
of \jackiw\ and note that, since the nonzero components of $\alpha$ generate fermion zero-modes, they
necessarily satisfy a non-trivial algebra of anti-commutation relations. The nonzero components of $\alpha$ 
must therefore be represented as operators acting on a space of quantum mechanical spin states \jackiw, which
in the present case is simply the BPS monopole supermultiplet. The monopole superpartner solution is then seen
to be operator valued. To get actual numerical values for the fields expectation values must be taken in 
specific BPS spin states.

Calculation of the different terms in the expansion \taylor\ is straightforward. At
first order, the only nonzero term is the variation of the fermion $\psi$ already given above in \fermionvar.
At second order, the variations $\delta^2S$ and $\delta^2A_k$ vanish and
we find only nonzero variations for $P$ and $A_0$ given by
\eqn\secondorder{\delta^2 A_0=-\delta^2P= -4i\left(\alpha^\dagger\gamma^k\alpha\right)D_k S.}
We will see below that these reduce to dipole fields in the long range limit. Interestingly, the third and
fourth order variations of all the fields turn out to vanish. In particular, the third order variation of
$\psi$ is found to be 
\eqn\thirdorder{
\delta^3\psi^a=8i\left(\alpha^\dagger\gamma^k\alpha\right)\left\{
\gamma^0\gamma^lD_lD_kS^a+e\gamma^0\epsilon^{abc}(D_kS^b)S^c\right\}
P_+\alpha,}
which vanishes because $P_+\alpha=0$ for the broken supersymmetries. The fourth order variations of the 
bosonic fields then vanish because they are each proportional to $\delta^3\psi$. Note, the vanishing of the
third and fourth order variations found here is in contrast with the results of \ae\ on $N=2$ black hole 
superpartners, for which these variations are nonzero.

\newsec{The Angular Momentum Operator}
The long range limits of the superpartner fields $A_0$ and $P$ turn out to be
related in a simple way 
to the angular momentum operators for the fermion zero-mode states, which are constructed in the
following way.  The fermionic fields $\psi^a$ may be expanded in the monopole background as
\eqn\zeromodes{
\psi^{a\rho}=-2(\gamma^k)^\rho{}_\sigma\alpha^\sigma D_kS^a + \hbox{nonzero-modes},}
where $\rho,\sigma$ are spinor indices and we have explicitly displayed only the zeromode part of the
expansion. Using the orthogonality of zero-modes and nonzero-modes, we can then express 
the spinorial parameters $\alpha^\lambda$ and
$\alpha^\dagger_\lambda$ as
\eqn\transforms{
\alpha^\lambda= +{1\over 2M}\int d^3x(\gamma^l)^\lambda{}_\rho\psi^{a\rho}D_lS^a,\qquad
\alpha^\dagger_\lambda= -{1\over 2M}\int d^3x\psi^{a\dagger}_\rho(\gamma^l)^\rho{}_\lambda D_lS^a,}
where $M=4\pi v/e$ is the mass of the monopole\foot{Here we have made use of the result
$\int d^3x\,\eta^{kl}(D_kS^a)D_lS^a = -M$}.
Making use of the canonical anti-commutation relation for the fermions 
$\left\{\psi^{a\sigma}(\vec x),\psi^{b\dagger}_\eta(\vec y)\right\}
=\delta^{ab}\delta^\sigma_\eta\delta(\vec x-\vec y)$, we arrive at anti-commutation relations
for $\alpha^\lambda$ and $\alpha^\dagger_\lambda$
\eqn\anticom{
\left\{\alpha^\rho,\alpha^\dagger_\lambda\right\}=+{1\over 4M}\delta^\rho_\lambda .}
Because of the projection condition $P_-\alpha=\alpha$ satisfied by the broken supersymmetries, 
the anti-commutation relations 
\anticom\ can be interpreted as the algebra of two sets of 
fermionic creation and annihilation operators, giving a total of four states, the states of the short BPS
supermultiplet. The fields of monopole superpartner solutions constructed above are operator
valued in this space of states.

It is now straightforward to check that the operators 
$J^{kl}= 2iM \left (\alpha^\dagger\gamma^{kl}\alpha\right)$ 
satisfy the angular momentum algebra 
$\left[J^{kl},J^{mn}\right]= i\left( \eta^{lm}J^{kn}-\eta^{ln}J^{km}
-\eta^{km}J^{nl}+\eta^{kn}J^{lm}\right)$, 
and hence generate rotations on the quantum mechanical zero-mode space of states. Alternatively, we can write 
down the angular momentum vector $J^k=-\half\epsilon_{klm}J^{lm}$, which after making use of the 
identity $\half\epsilon_{klm}\gamma^{lm}=\gamma^k\Gamma_5$, is given by
\eqn\angmom{
J^k=2iM\left (\alpha^\dagger\gamma^{k}\alpha\right).}

\newsec{The Electric Dipole Moment}
We now turn to the long range limit of the electric field for the monopole superpartner solution.
In order to extract this from the expression \secondorder\ for
$\delta^2A_0^a$, 
we need to plug in the long-distance limits of the fields of the zeroth order monopole solution,
which are given by
\eqn\longdistance{
S^a= \left({v\over r}-{1\over er^2}\right)x^a,\qquad
A^a_i=\epsilon_{aij}{x^j\over er^2},}
Far from the monopole core, we then have
\eqn\longgauge{A_0^a={1\over 2}\delta^2 A_0^a= -2i{x^ax^k\over er^4}
\left(\alpha^\dagger\gamma^k\alpha\right).}
We still need to compute the non-abelian field strength from \longgauge\ and extract the $U(1)$ part.
The result for the long range abelian electric field obtained in this way is
\eqn\dipolefield{E^i=
F_{0i}\equiv {1\over v}S^aF_{0i}^a=-{2i\over e}\left(\alpha^\dagger\gamma^{k}\alpha\right)
\left\{{3x^kx^i\over r^5}-{\delta^{ki}\over r^3}\right\} ,}
which is a dipole field with dipole moment vector $\vec p= -{2i\over e}(\alpha^\dagger\vec\gamma\alpha)$.
The electric dipole moment $\vec p$ is clearly proportional to the zero-mode angular momentum operator 
$\vec J$ in \angmom. If we then define a gyroelectric
ratio $g$ for the monopole superpartners via the relation 
\eqn\ratio{\vec p= - (gQ_m/8\pi M)\vec J,}
where $Q_m=4\pi/e $ is the
magnetic charge of the monopole, we find $g=2$ for the
monopole superpartner solution in agreement with general results for a short $N=2$ multiplet 
\porrati,\liu. The minus sign in the relation \ratio\ corresponds to the minus sign in the
electromagnetic duality relation 
\eqn\duality{\vec E\longrightarrow\vec B,\qquad\vec B\longrightarrow -\vec E.}
A current loop of magnetic monopoles has an electric dipole moment which points opposite to the
magnetic dipole moment of an electric current loop. We note finally that the vanishing of the fourth order
variations of the bosonic fields implies a vanishing quadrupole moment tensor for all states in
the monopole BPS multiplet.

\bigskip\noindent
{\bf Note Added:} After this work was completed, a much earlier derivation \osborn\ of the result $g=2$, for
the electric dipole moment of a BPS monopole, was brought to our attention. The result in \osborn\ was obtained
by considering the change in energy of a monopole in a weak external electric field. Hence, the present, very
different calculation can be considered as offering a complementary perspective.

\bigskip\noindent
{\bf Acknowledgements: }  We thank Jerome Gauntlett, Jeff Harvey and Jennie Traschen for helpful discussions
and correspondence.

\listrefs
\end

It is well known that magnetic monopoles satisfying the Dirac quantization condition 
are necessarily the finite configurations with non-zero topological charge in the Yang-Mills-Higgs theory,whose
Lagrangian is 
\eqn\lagrangian{
{\cal L}=-{1\over4}F_{\mu\nu}^a F^{a\mu\nu} +{1\over2}D_{\mu}\Phi^aD_{\mu}\Phi^a-V(\Phi), }
where
\eqn\fields{\eqalign{
F_{\mu\nu}^a&=\partial_{\mu}A_{\nu}^a-\partial_{\nu}A_{\mu}^a +e\epsilon^{abc}A_{\mu}^b A_{\nu}^c,\cr
D_{\mu}\Phi^a&=\partial_{\mu}\Phi^a +e\epsilon^{abc}A_{\mu}^b \Phi^c,\cr}
}
with $a,b,c=1,2,3$ labeling the adjoint representation of $SU(2)$. 
For non-zero vaccuum expectation value of $\Phi$ 
\eqn\potential{
V(\Phi)=\lambda(\Phi^a\Phi^a-v^2)^2/4.
}
The equations of motion are
\eqn\motion{\eqalign{
D_{\mu}F^{\mu\nu}&=e\epsilon^{abc}\Phi^b D^{\nu}\Phi^c, \cr
(D^{\mu}D_{\mu}\Phi)^a&=-\lambda\Phi^a(\Phi^b\Phi^b-v^2), \cr} }
and the Bianchi identity
\eqn\bianchi{
D_{\mu}\ast F^{a\mu\nu}=0.
}

The energy density of field configuration is 
\eqn\energy{
\Theta_{00}={1\over 2}(\vec E^a\cdot\vec E^a +\vec B^a\cdot\vec B^a+\Pi^a\Pi^a
+\vec D\Phi^a\cdot\vec D\Phi^a)+V(\Phi), }
where $\Pi^a$ is the conjugate momentum of field $\Phi$,$\Pi^a=D_0\Phi^a$ and 
non-abelian electric and magnetic fields are 
\eqn\nonabelian{E^{ai}=-F^{a0i},\qquad B^{ai}=-{1\over2}\epsilon^{ijk}F_{jk}^a. }
The vacuum is given by a configuration with vanishing gauge field, with a constant Higgs filed $\Phi^a$
\eqn\vacuum{
F^{a\mu\nu}=D^\mu \Phi^a=V(\Phi)=0,\qquad \Phi^a\Phi^a=v^2.}
A constant Higgs field breaks $SU(2)$ gauge symmetry to $U(1)$ subgroup.

Magnetic charge is defined as
\eqn\magnetic{
q={1\over4\pi}\int_{S_{\infty}^2}\vec B\cdot d\vec S ={1\over4\pi v}
\int_{S_{\infty}^2}\Phi^a\vec B^a\cdot d\vec S ={1\over4\pi v}\int\vec B^a\cdot (\vec D\Phi)^a d^3r, 
}
using the Bianchi identity and integration by parts. 
The energy of the static configuration with vanishing electric field is given by
\eqn\static{\eqalign{
M_M&=\int d^3r\left({1\over2}(\vec B^a\cdot\vec B^a +\vec D\Phi^a\cdot\vec D\Phi^a)+V(\Phi)\right)\cr
&\geq{1\over2}\int d^3r(\vec B^a-\vec D\Phi^a)\cdot (\vec B^a-\vec D\Phi^a)+4\pi vq,\cr}
}
using \magnetic. From this we have Bogomol'nyi bound 
\eqn\bogo{M_M\geq 4\pi vq. }
and two conditions
\eqn\conditions{V(\Phi)=0,\qquad \vec B^a=\vec D\Phi^a, }
which saturate the bound. The second condition is called as Bogomol'nyi equation. 
In supersymmetric theories which have potentials with exact flat directions protected by supersymmetry, symmetry
breaking makes sense with the first condition if we impose a boundary condition 
\eqn\boundary{ \Phi^a\Phi^a\to v^2\;\;\;{\rm as}\;\, r\to\infty, } 
for arbitrary v.

A simple solution to the equation of motion with Bogomol'nyi condition is 
found by Prasad-Sommerfield in terms of two radial functions $H,\,K$
\eqn\prasad{\eqalign{
\Phi^a&={\hat r^a\over er}H(ver),\cr
A_i^a&=-\epsilon_{ij}^a{{\hat r}^j\over er}(1-K(ver)).\cr} }
with the boundary conditions
\eqn\conditions{\eqalign{
K(ver)&\rightarrow 1,\qquad H(ver)\rightarrow 0,\quad {\rm as}\;r\rightarrow 0,\cr 
K(ver)&\rightarrow  0,\qquad H(ver)/(ver)\rightarrow 1,\quad {\rm as}\quad 
r\rightarrow \infty.\cr }}
where
\eqn\functions{
H(y)=y\coth y-1, \qquad
K(y)={y\over\sinh y}.}
For large $r$ we thus have
\eqn\asymptotic{
\Phi^a\rightarrow v\hat r^a-{\hat r^a\over er} A_i^a\rightarrow -\epsilon_{ij}^a{{\hat r}^j\over er}.} 

\newsec{Monopoles in N=2 Supersymmetric Gauge Theory} 
The Lagrangian of $N=2$ Super Yang-Mills theory is given by 
\eqn\super{\eqalign{
{\cal L}_{N=2}&={\rm Tr}(-{1\over4}F_{\mu\nu}F^{\mu\nu}
-{1\over4}(D_{\mu}P)^2-{1\over2}(D_{\mu}S)^2-{e^2\over2}[S,P]^2 \cr &
+i\bar\psi\gamma^{\mu}D_{\mu}\psi-e\bar\psi[S,\psi] -e\bar\psi\gamma_5[P,\psi]) \cr} }
where all fields
are elements of $SU(2)$ Lie algebra, {\it i.e.} $S=S^a T^a$ {\it etc.} and $S$ and $P$ are scalar Higgs fields.
This Lagrangian is invariant under the supersymmetry transformations 

\eqn\susyvariations{\eqalign{
\delta A_{\mu}&=i\bar\alpha\gamma_{\mu}\psi -i\bar\psi\gamma_{\mu}\alpha, \cr
\delta P&=\bar\alpha\gamma_5\psi
-\bar\psi\gamma_5\alpha, \cr
\delta S&=i\bar\alpha\psi
-i\bar\psi\alpha, \cr
\delta\psi&=(\sigma^{\mu\nu}F_{\mu\nu}-i\gamma^\mu D_\mu S +\gamma^\mu D_\mu P\gamma_5-i[P,S]\gamma_5)\alpha, \cr }}
with $\alpha$ the supersymmetry parameter of the grassmann valued Dirac spinor 
which is equivalently two Majorana spinors. 

We are now going to find the electric dipole moment of magnetic monopole doing higher order 
supersymmetry variation of fields. All the fields are considered as
\eqn\taylorseries{
X_\mu=X_{\mu}^{(0)}+X_{\mu}^{(1)}+{1\over2}X_{\mu}^{(2)}+\dots }
where $X_{\mu}^{(n)}=\delta\,.\,.\,.\,\delta X_{\mu}$ with $n\,\delta$s. 
We start with a classical solution with the fermion field set to zero,choose gauge condition $A_0=0$ and set
$P^a=0$ by chiral rotation. Our boundary condition is as before $S^a S^a \rightarrow v^2$ as $r\rightarrow\infty$
and the Bogomol'nyi equation becomes 
\eqn\susybogo{
B_i=D_i S =-{1\over2}\epsilon_{ijk}F^{jk}. }
Then we have
\eqn\firstorder{\eqalign{
\psi^{(1)}&=\delta\psi=(\sigma^{\mu\nu}F_{\mu\nu}-\rlap/{D}S)\alpha\cr 
&=-\gamma^i B_i(1-\Gamma_5)\alpha\cr
&=-2(\gamma^i D_iS)\alpha_-. \cr} }
Our conventions and definitions used above are 
\eqn\conventions{
\epsilon^{0123}=+1,\qquad \eta^{\mu\nu}=(1,-1,-1,-1), }
and
\eqn\moreconventions{\eqalign{
\{\gamma^{\mu},\gamma^{\nu}\}&=2g^{\mu\nu},\qquad \sigma^{\mu\nu}={1\over4}[\gamma^{\mu},\gamma^{\nu}],\cr
\gamma_5&=i\gamma^0\gamma^1\gamma^3\gamma^3,\qquad \Gamma_5=i\gamma^0\gamma_5 \cr
\Gamma_{\pm}&={1\over2}(1\pm\Gamma_5),\qquad \alpha_{\pm}=\Gamma_{\pm}\alpha.\cr} }
where the supersymmetries $\alpha_+$ are unbroken in the monopole background while the $\alpha_-$ 
supersymmetries are broken,that is $\alpha_+=0$.
Equation \firstorder\ for the broken supersymmetries give zero energy Grassmann 
variations of the monopole solution.
Thus we have
\eqn\allfirst{\eqalign{
\bar\psi^{(1)}&=-2\bar\alpha_-(\gamma^i D_i S),\cr
 A_{\mu}^{(1)}&=0,\cr
P^{(1)}&=0,\cr
S^{(1)}&=0\cr}}
The second order variations are by using the relation 
\eqn\trick{
\gamma_5\alpha_-=i\gamma^0\alpha_-,
}
\eqn\secondorder{\eqalign{
A_{0}^{(2)}&=P^{(2)}
=-4i(D_i S)(\bar\alpha_-\gamma_0\gamma^i\alpha_-), \cr 
A_{i}^{(2)}&=0=S^{(2)}=\psi^{(2)}=\bar\psi^{(2)}. \cr}}
Unexpectedly all the higher order variations are identically zero. 

Now we can write the component of the electric field arising from moving magnetic charge of monopole as 
\eqn\electric{
E^{ia}=\partial_i{A_0^{(2)a}\over2}
-e\epsilon^{abc}{A_0^{(2)b}\over2}A_i^{(0)c}, }
and by using equation \asymptotic\ and the finite energy condition 
\eqn\abelianpart{
E_i=E_i^a{S^a\over v},
}
with
\eqn\anothertrick{
-i\gamma_i\gamma_0\gamma_5={1\over4}\epsilon_{ijk}\gamma^{jk},\qquad \gamma_{kl}\equiv [\gamma_k,\gamma_l],
}
we have
\eqn\finally{\eqalign{
E_i&={i\over2}\epsilon^{jkl}(\alpha_-^{\dagger}\gamma_{kl}\alpha_-) \left({1\over
ver}-1\right)\left({\eta_{ij}\over e r^3} -{3x_i x_j\over e r^5}\right)\cr 
&=\left({3x_i x_j\over
r^5}-{\eta_{ij}\over e r^3}\right) {\cal P}^j+\dots\cr} }
where
\eqn\operator{
{\cal P}^j={i\over2e}\epsilon^{jkl}\alpha_-^{\dagger}\gamma_{kl}\alpha_-, }
denotes the electric dipole moment of moving magnetic monopole.
Since the angular momentum of spin ${1\over2}$ fermion is defined as 

\eqn\angmom{
J^{\mu\alpha\beta}=x^\alpha\Theta^{\mu\beta} -x^\beta\Theta^{\mu\alpha}
+{1\over2}\bar\psi\{\gamma^{\mu},{i\over 4}\gamma^{\alpha\beta}\}\psi, }
where $\Theta^{\mu\nu}$ is the energy-momentum tensor density. 
The purely spin part which is arising from the last term is written 

\eqn\spinangmom{
J^{0jk}={1\over2}\bar\psi^{(1)}
\{\gamma^{0},{i\over4}\gamma^{\alpha\beta}\}\psi^{(1)}, }

and thus
\eqn\result{\eqalign{
\epsilon_{ijk}J^{0jk}&=-{i\over2}\epsilon_{ijk}\alpha_-^{\dagger} \gamma^{jk}\alpha_-D^l S D_l S\cr
&=-e{\cal P}^i D^l S D_l S,\cr
}}
is straight forward.
Therefore spin component ${\cal S}^i$ is 

\eqn\spinvector{\eqalign{
{\cal S}^i&=\int{\rm d}^3x\epsilon^{ijk}J_{0jk}\cr
&=-e{\cal P}^i\int{\rm d}^3x\,\partial_l(S^a D^l S^a)\cr
&=-e{\cal P}^i\int{\rm d}\theta{\rm d}\varphi\,{\rm sin}\theta\, r x_l(S^a D^l S^a)\cr 
&=-4\pi e{\cal P}^i\left[-{r\over e}({v\over r}-{1\over er^2}) \right]_{r\rightarrow\infty}\cr 
&=4\pi v{\cal P}^i \cr}}
where we use equations \motion\ and \asymptotic\ and

\eqn\yetanothertrick{
D_l(S^a D^l S^a)=\partial_l(S^a D^l S^a), }
hence the electric dipole moment of monopole ${\cal P}^i$ is 

\eqn\dipolemoment{
{\cal P}^i={{\cal S}^i\over4\pi v}\equiv{g_e q\over2M_M}{\cal S}^i,}
where $g_e$ is the gyroelectric ratio and finally using equation \bogo\ we have

\eqn\gyroelectric{
g_e=2.
}

\newsec{The Electric Dipole Moment}

\listrefs
\end

\newcommand{\eqn{} {\eqn{gin{equation}}
\newcommand{}} {\end{equation}}
\newcommand{\eqn{a} {\eqn{gin{eqnarray}}
\newcommand{}a} {\end{eqnarray}}

\newcommand{\ba} {\eqn{gin{array}}
\newcommand{\ea} {\end{array}}
\newcommand{\cr} {\noindent}
\newcommand{\cr} {\nonumber \\}
\newcommand{\no} {\nonumber}
\newcommand{\ul} {\underline}
\newcommand{\wt} {\widetilde}
\newcommand{\wh} {\widehat}
\newcommand{\bfl} {\eqn{gin{flushleft}}
\newcommand{\efl} {\end{flushleft}}